%% file: ANA-EXOT-2023-16-PAPER.tex
\documentclass[cernpreprint=true, atlasdraft=false, USenglish, orcidlogo, texmf, orcidlogo]{atlasdoc}
 
\usepackage[subfigure,rotating]{atlaspackage}
\usepackage{atlasbiblatex}
\usepackage{multirow}
 
\usepackage{atlasphysics}
 
\usepackage{ANA-EXOT-2023-16-PAPER-defs}

\graphicspath{{logos/}{figures/}}
 
\usepackage{ANA-EXOT-2023-16-PAPER-defs}

\AtlasTitle{Search for heavy Majorana neutrinos in $e^{\pm} e^{\pm}$ and $e^{\pm} \mu^{\pm}$ final states via WW scattering in $pp$ collisions at $\sqrt{s}=13$ TeV with the ATLAS detector}

\AtlasAbstract{
A search for heavy Majorana neutrinos in scattering of same-sign $W$ boson pairs in proton--proton collisions at  $\sqrt{s}= 13$~TeV at the LHC is reported. The dataset used corresponds to an integrated luminosity of 140~fb$^{-1}$, collected with the ATLAS detector during 2015--2018. The search is performed in final states including a same-sign $ee$ or $e\mu$ pair and at least two jets with large invariant mass and a large rapidity difference.  No significant excess of events with respect to the Standard Model background predictions is observed.  The results are interpreted in a benchmark scenario of the Phenomenological Type-I Seesaw model. New constraints are set on the values of the $\vert V_{e N} \vert^{2}$  and $\vert V_{e N} V^{*}_{\mu N} \vert$  parameters for heavy Majorana neutrino masses  between 50~GeV and 20~TeV, where $V_{\ell N}$ is the matrix element describing the mixing of the heavy Majorana neutrino mass eigenstate with the Standard Model neutrino of flavour $\ell = e, \mu$.  The sensitivity to the Weinberg operator is investigated and constraints on the effective $ee$  and $e\mu$ Majorana neutrino masses are reported. The statistical combination of the $ee$ and $e\mu$ channels with the previously published $\mu\mu$ channel is performed.}

\AtlasRefCode{EXOT-2023-16}
 
\PreprintIdNumber{CERN-EP-2024-083}

\AtlasJournal{Phys.\ Lett.\ B.}
\AtlasJournalRef{Phys.\ Lett.\ B 856 (2024) 138865}
\AtlasDOI{10.1016/j.physletb.2024.138865}

\hypersetup{pdftitle={ATLAS document},pdfauthor={The ATLAS Collaboration}}
 
\begin{document}
 
\maketitle
 
\tableofcontents

\section{Introduction}
 
The observation of neutrino oscillations~\cite{Super-Kamiokande:2005mbp, SNO:2002tuh} has conclusively demonstrated that neutrinos have nonzero masses.  Stringent bounds on the neutrino masses have been obtained from cosmological observations~\cite{Planck:2018vyg, eBOSS:2020yzd} and from direct experimental measurements such as the measurements of kinematics of nuclear $\beta$-decays of tritium~\cite{KATRIN:2021uub}. Massive neutrinos can be incorporated into the Standard Model (SM) by introducing additional Dirac mass terms with right-handed chiral neutrino states.  The smallness of neutrino masses suggests that another mechanism of generation of neutrino masses might be present via the inclusion of Majorana mass terms for left- and right-handed chiral neutrino states.
 
One of the most compelling scenarios is the hypothesized existence of Majorana mass terms for neutrinos~\cite{Atre:2009rg,Deppisch:2015qwa,Cai:2017jrq,Cai:2017mow}. The Majorana nature of neutrinos can be probed by searching for neutrinoless double beta decays characterized by the appearance of two same-charge (same-sign) electrons with an absence of neutrinos in the final state~\cite{Dolinski:2019nrj}.  Majorana neutrino masses are present in beyond the SM (BSM) theories that include new, heavy leptons or extended scalar sectors, collectively known as Seesaw mechanism, or in theories with extended gauge sectors (e.g. Left-Right Symmetric Model, Grand Unified Theories)~\cite{Ma:1998dn}.  In the context of the Type-I Seesaw mechanism, the small neutrino masses $m_{\nu}$ are explained by introducing a new heavy Majorana neutrino mass eigenstate ($N$) with $m_{\nu} \approx \mathcal{O} (\upsilon^2 / m_N)$, where $\upsilon$ is the Higgs vacuum expectation value $\upsilon\approx246~\GeV$ and $m_{N}$ is the mass of the heavy Majorana neutrino.
 
This letter presents the first search of heavy Majorana neutrinos in scattering of same-sign $W$ boson pairs in the same-sign dielectron ($ee$) or  lepton flavour violating electron-muon ($e\mu$) final states at the Large Hadron Collider (LHC) using proton--proton (\pp) collisions at $\sqrt{s}=13~\TeV$. The dataset corresponds to an integrated luminosity of $140.1 \pm 1.2$~fb$^{-1}$~\cite{DAPR-2021-01,LUCID2}, collected with the ATLAS detector during 2015--2018.  Searches for heavy Majorana neutrinos in the $W^{\pm}W^{\pm}$ scattering topology using the same-sign dimuon ($\mu\mu$) final state were reported  by the ATLAS~\cite{EXOT-2020-06} and CMS~\cite{CMS:2022hvh} Collaborations. These searches extend the reach for heavy Majorana neutrino masses with respect to the searches via resonant production of a charged lepton and $N$~\cite{ATLAS:2019kpx,ATLAS:2015gtp,CMS-EXO-17-028,CMS:2018iaf} at the LHC as the resonant production channels become kinematically inaccessible beyond the \TeV~scale~\cite{Fuks:2020att}. Searches for heavy Majorana neutrinos lighter than the $Z$ boson were conducted at LEP~\cite{L3:1992xaz, DELPHI:1996qcc, L3:2001zfe,}.  Searches for light Majorana neutrinos with $m_N < 50~\GeV$ were reported by the LHCb Collaboration~\cite{LHCb:2014osd}.
 
The results are interpreted in the context of Phenomenological Type-I Seesaw model~\cite{delAguila:2008cj,Atre:2009rg}, where heavy Majorana neutrinos couple to SM particles through mass-mixing with SM neutrinos. The mixing is parameterized with a complex-valued matrix element $V_{\ell N}$, where $\ell = e, \mu$.  The two free parameters in this phenomenological approach are $V_{\ell N}$ and $m_{N}$. For simplicity, the benchmark scenario is considered without $\tau$ flavour mixing and only the contribution of the lightest mass eigenstate $N$ to the amplitude is included.  Results are also interpreted in an effective field theory where the Majorana masses can be generated most minimally with a dimension-5 Weinberg operator~\cite{Weinberg:1979sa}.  The effective Majorana mass is given by $\vert  m_{\ell\ell'} \vert = \vert C_5^{\ell\ell'} \vert \upsilon^2/\Lambda$, where $\Lambda$ is the scale at which the particles mediating the BSM signal become relevant degrees of freedom and $C_5^{\ell\ell'}$ is a flavour-dependent Wilson matrix coefficient, which for the  $\ell \ell' = ee$ reduces to the effective mass in neutrinoless double beta decay experiments~\cite{Dolinski:2019nrj, Fuks:2020zbm}.  Neutrinoless double beta decay nuclear experiments cannot probe  $\vert  m_{e\mu} \vert$ and $\vert  m_{\mu\mu} \vert$ as the production of a muon is kinematically forbidden.

\begin{figure}[t!bp]
\centering
\subfigure[]{\includegraphics[width=0.30\textwidth]{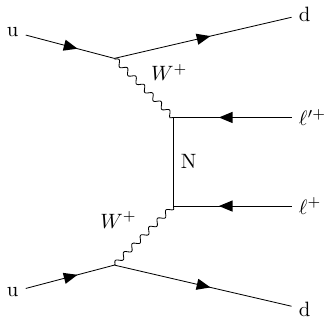}}
\subfigure[]{\includegraphics[width=0.30\textwidth]{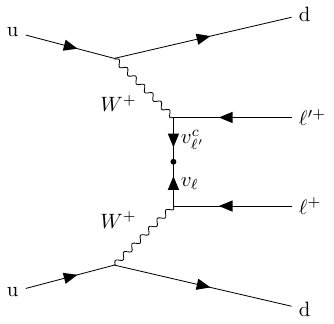}}
\caption{Representative Feynman diagrams of same-sign $\ell^{+}\ell^{'+}$ production in $W^{+} W^{+}$ scattering mediated by (a) a Majorana neutrino $N$ and (b) the dimension-5 Weinberg operator. The corresponding diagrams with negatively--charged leptons are also considered.}
\label{fig:feyn_sig}
\end{figure}

Figure~\ref{fig:feyn_sig} shows representative Feynman diagrams of same-sign $\ell^{+}\ell^{'+}$ production in $W^{+} W^{+}$ scattering mediated by a Majorana neutrino $N$ and mediated by the dimension-5 Weinberg operator. The production cross section in the Phenomenological Type-I Seesaw model is proportional to  $\vert V_{e N} \vert^4 $ and $\vert V_{e N} V^{*}_{\mu N} \vert^2 $ for the $ee$ and $e\mu$ final states, respectively. The production cross section in the dimension-5 Weinberg operator model is proportional to  $ \vert C_5^{\ell\ell'}  \vert^2 /\Lambda^2$.  Compared to the dimension-5 Weinberg operator, the signal production in the Phenomenological Type-I Seesaw model is a dimension-7 process~\cite{Fuks:2020att,Fuks:2020zbm, Dicus:1991fk}.

Signal candidate events contain exactly two identified same-sign leptons  ($ee$ or $e\mu$) and two jets with a large rapidity separation and a high dijet mass. The signal processes contain no neutrinos, and thus no significant missing transverse momentum in the final state. The requirements on the dijet mass and rapidity separation increase the sensitivity to the vector boson scattering topology and reduce the contribution from the SM production of heavy gauge boson pairs mediated by the strong interaction. The dominant SM background processes are $WZ$ and electroweak (EW) same-sign $WW$  production in association with two jets, where the bosons decay leptonically. These background processes are estimated with Monte Carlo (MC) simulated events and their modelling is constrained in dedicated signal-depleted control regions (CRs).  Data-driven techniques assisted by MC simulation are used to estimate backgrounds including electrons or muons not originating from the prompt decay of particles such as $W$ or $Z$ bosons (referred to as non-prompt leptons) and backgrounds including electron charge misidentification.  Other minor backgrounds, including contributions mainly from the $\ell\gamma jj$, $ZZ$ and \tZq background processes, are estimated using MC simulation.

\section{ATLAS detector}

The ATLAS experiment~\cite{PERF-2007-01} at the LHC is a multipurpose particle detector
with a forward--backward symmetric cylindrical geometry and a near \(4\pi\) coverage in
solid angle.\footnote{ATLAS uses a right-handed coordinate system with its origin at the nominal interaction point (IP) in the centre of the detector and the \(z\)-axis along the beam pipe. The \(x\)-axis points from the IP to the centre of the LHC ring, and the \(y\)-axis points upwards. Cylindrical coordinates \((r,\phi)\) are used in the transverse plane, \(\phi\) being the azimuthal angle around the \(z\)-axis. The pseudorapidity is defined in terms of the polar angle \(\theta\) as \(\eta = -\ln \tan(\theta/2)\). The rapidity is defined as $y=\frac{1}{2} \ln{\frac{E+p_z}{E-p_z}}$, where $E$ is the energy and $p_z$ is the longitudinal component of the momentum.
}
It consists of an inner tracking detector (ID) surrounded by a thin superconducting solenoid
providing a \qty{2}{\tesla} axial magnetic field, electromagnetic and hadronic calorimeters, and a muon spectrometer (MS).
The inner tracking detector covers the pseudorapidity range \(|\eta| < 2.5\).
It consists of silicon pixel, silicon microstrip, and transition radiation tracking detectors.
Lead/liquid-argon (LAr) sampling calorimeters provide electromagnetic (EM) energy measurements
with high granularity within the region \(|\eta|< 3.2\).
A steel/scintillator-tile hadronic calorimeter covers the central pseudorapidity range (\(|\eta| < 1.7\)).
The endcap and forward regions are instrumented with LAr calorimeters
for EM and hadronic energy measurements up to \(|\eta| = 4.9\).
The MS surrounds the calorimeters and is based on
three large superconducting air-core toroidal magnets with eight coils each.
The field integral of the toroids ranges between \num{2.0} and \qty{6.0}{\tesla\metre}
across most of the detector.
The MS includes a system of precision tracking chambers up to \(|\eta| = 2.7\) and fast detectors for triggering up to \(|\eta| = 2.4\).
The luminosity is measured mainly by the LUCID--2~\cite{LUCID2} detector, which is located close to the beampipe.
A two-level trigger system is used to select events~\cite{TRIG-2016-01}.
The first-level trigger is implemented in hardware and uses a subset of the detector information
to accept events at a rate below \qty{100}{\kHz}.
This is followed by a software-based trigger that
reduces the accepted event rate to \qty{1}{\kHz} on average
depending on the data-taking conditions.
A software suite~\cite{ATL-SOFT-PUB-2021-001} is used in data simulation, in the reconstruction
and analysis of real and simulated data, in detector operations, and in the trigger and data acquisition
systems of the experiment.

\section{Event simulation}
 
MC event simulations were produced using the ATLAS event simulation infrastructure~\cite{SOFT-2010-01} and \textsc{Geant4}~\cite{Agostinelli:2002hh}. The effect of additional $\pp$ interactions per bunch crossing (\pileup) is accounted for by overlaying the hard-scattering process with minimum-bias events generated with \PYTHIA[8.186]~\cite{Sjostrand:2007gs} using the \NNPDF[2.3lo] set of parton distribution functions (PDF)~\cite{Ball:2012cx} and the A3 set of tuned parameters~\cite{ATL-PHYS-PUB-2016-017}. Different \pileup conditions between data and simulation are taken into account by reweighting the mean number of interactions per bunch crossing in simulation to the number observed in data. The \NNPDF[3.0nlo]~\cite{Ball:2014uwa} PDF set was used in all matrix element calculations unless stated otherwise. All
simulated samples were processed through the same reconstruction algorithms and analysis chain as the data. For all samples of simulated signal events, the parton shower, hadronisation, and underlying-event modelling by \PYTHIA used the A14 tune~\cite{ATL-PHYS-PUB-2014-021} and the \NNPDF[2.3lo] PDF set. The \EVTGEN[1.7.0] program~\cite{Lange:2001uf} was used to model the decays of bottom and charm hadrons. All MC samples were normalized to the highest-order theory predictions available.

Signal samples corresponding to the Phenomenological Type-I Seesaw model were simulated following the prescription given in Ref.~\cite{Fuks:2020att}.
The samples were produced using the \MGNLO[2.7.2]~\cite{Alwall:2014hca} generator at next-to-leading-order (NLO) in QCD after importing the default variant of the \textsc{HeavyN}~\cite{Degrande:2016aje, Alva:2014gxa} \textsc{FeynRules} UFO libraries~\cite{Christensen:2009jx, Alloul:2013bka}. The \NNPDF[3.1nlo] PDF set was used. All \textsc{HeavyN} events were processed through \PYTHIA[8.243]~\cite{Sjostrand:2014zea}. Twenty mass points in the range $m_{N} = 50$\,\GeV~to 25\,\TeV were generated.  The signal sample for the Weinberg operator was simulated by using the prescription given in Ref.~\cite{Fuks:2020zbm}. For the hard parton-level scattering processes, \MGNLO[2.9.3] was used at NLO in QCD after importing the default variant of the \textsc{SMWeinberg}~\cite{Fuks:2020zbm} \textsc{FeynRules} UFO libraries. The NNPDF3.1luxQED~\cite{Ball:2014uwa} PDF set was used.
All \textsc{SMWeinberg} events were processed through \PYTHIA[8.245].
 
Background samples corresponding to EW same-sign $WW$ production and decay were simulated with diagrams including exactly six orders of the EW coupling~\cite{ATL-PHYS-PUB-2019-004}.  The simulation of the strong production processes and the interference between EW and strong contributions includes diagrams with exactly four and five EW vertices~\cite{ATL-PHYS-PUB-2017-005}, respectively. All SM same-sign $WW$ contributions were simulated with \MGNLO[2.6.7] at leading order (LO), which was interfaced to the \PYTHIA[8.244]~\cite{Sjostrand:2014zea} parton shower model.  More details of the generator settings used to obtain a better description of deep-inelastic scattering data~\cite{Cabouat:2017rzi} are given in Ref.~\cite{ATL-PHYS-PUB-2019-004}.
 
Samples for diboson ($VV$, $V=W/Z$) production were simulated with the
\SHERPA[2.2.1] or 2.2.2~\cite{Bothmann:2019yzt} generator depending on the process using the \NNPDF[3.0nnlo] PDF set.
Fully leptonic final states and semileptonic final states, where one boson
decays leptonically and the other hadronically, were generated using
matrix elements with NLO accuracy in QCD for up to one additional parton
and with LO accuracy for up to three additional parton
emissions. Samples for the loop-induced processes $gg \to VV$ were
generated using LO matrix elements for up to one
additional parton emission.
The production of triboson ($VVV$) events was simulated with the
\SHERPA[2.2.2] generator, accurate to NLO in QCD for the inclusive process and to LO for up to
two additional parton emissions, using factorized gauge-boson decays.
The matrix element calculations for individual $VV$ and $VVV$ processes were matched
and merged with the \SHERPA parton shower based on Catani--Seymour
dipole factorization~\cite{Gleisberg:2008fv,Schumann:2007mg} using the MEPS@NLO
prescription~\cite{Hoeche:2011fd,Hoeche:2012yf,Catani:2001cc,Hoeche:2009rj}.
The virtual QCD corrections for matrix elements with NLO accuracy were
provided by the \OPENLOOPS[1] library~\cite{Buccioni:2019sur,Cascioli:2011va,Denner:2016kdg}.
 
Samples for $V\gamma$ processes~($V=W,Z/\gamma^*$) were simulated with  \SHERPA[2.2.11] using the \NNPDF[3.0nnlo] PDF set. All off-shell contributions were taken into account. The NLO matrix elements with up to one additional parton and LO matrix elements with up to three partons were merged with the parton shower using an MEPS@NLO merging scale. The photon was required to be isolated from leptons~\cite{Frixione:1998jh}. The EW production of $V\gamma$ was modelled at LO using \MGNLO[2.6.5] and \PYTHIA[8.240].
 
Additional samples were used to simulate minor backgrounds.
The production of \ttV , $t\bar{t}WW$, \tZq\ and \tWZ\ events was modelled using the
\MGNLO[2.3.3] generator at NLO, processed through \PYTHIA[8.210]. 
The production of $t\bar{t}$ and single-top-quark events was simulated using the
\POWHEGBOX[v2]~\cite{Frederix:2012dh,Nason:2004rx,Frixione:2007vw,Alioli:2010xd}
generator at NLO in QCD, interfaced to \PYTHIA[8.230], and
the production of $V+$jets events~\cite{PMGR-2021-01} was simulated with the \SHERPA[2.2.11] generator.

\section{Object reconstruction and event selection}

A set of single-electron~\cite{TRIG-2018-05} and single-muon triggers~\cite{TRIG-2018-01} is used in the $e\mu$ channel. The transverse momentum ($\pT$) thresholds are in the range $20$--$26~\GeV$ depending on the lepton flavour and data-taking period.  The $ee$ channel uses dielectron triggers with \pt thresholds of 12~\GeV, 17~\GeV and 24~\GeV for the 2015, 2016 and 2017 to 2018 data periods, respectively. All detector subsystems were required to be operational during data taking and to satisfy data quality requirements~\cite{DAPR-2018-01}.
 
Events are required to have at least one collision vertex reconstructed from at least two ID tracks with
\(\pt > 500~\MeV\). For events with several collision vertices, the one with the largest sum of the squared
transverse momenta of the associated tracks is taken as the hard-scatter vertex~\cite{ATL-PHYS-PUB-2015-026}.
 
Electrons are reconstructed from isolated electromagnetic calorimeter clusters, which are matched to tracks in the ID~\cite{EGAM-2018-01}. Baseline electrons are required to satisfy a `Loose' likelihood-based identification criterion. Further requirements on the longitudinal and transverse impact parameters are imposed. The transverse impact parameter significance\footnote{The transverse impact parameter significance is defined as $|d_0|/\sigma(d_0)$, where $d_0$ is the distance of closest approach of the $e$ or $\mu$ track to the primary vertex in the transverse plane and $\sigma(d_0)$ is its uncertainty.} is required to satisfy ${|\dzero|/\sigma(\dzero)} < 5$.  The longitudinal impact parameter\footnote{The longitudinal impact parameter is equal to $|z_0 \sin \theta|$, where $z_0$ is the difference between the value of the $z$ coordinate of the point on the track at which $d_0$ is defined as the longitudinal position of the primary vertex.} is required to satisfy ${|\zzsth|<0.5}$~mm.  Baseline electrons are also required to have $\pt >4.5~\GeV$ and $|\eta |<$ 2.47.  Signal electrons have the same requirements as baseline electrons, satisfying $\pt>27~\GeV$ and $\eta<$ 2.47, with electrons in the transition region 1.37 $<|\eta|<$1.52 removed. In addition signal electrons must satisfy the `Tight' likelihood-based identification and `Gradient' isolation requirement~\cite{EGAM-2018-01}.  A charge-selector tool uses shower shape and track-to-cluster matching variables to reject electron candidates where the charge is likely misidentified~\cite{EGAM-2018-01}. \enquote{Background} electrons with $\pt>27~\GeV$, used to estimate the background processes with non-prompt electrons, are required to pass Medium likelihood-based identification~\cite{EGAM-2018-01}, with no requirements on the isolation criteria. Background electrons are required to fail the signal electron selection to ensure that the samples of signal and background electrons are statistically independent.

Muons  are reconstructed~\cite{MUON-2018-03} from tracks in the MS and matched to a corresponding track in the ID where possible. Baseline muons are required to have $\pT>3~\GeV$ and $|\eta|<2.7$. Further selections on the longitudinal and transverse impact parameters are imposed by requiring  ${|\dzero|/\sigma(\dzero)} < 15$ and ${|\zzsth|<1.5}$~mm.  Baseline muon candidates must satisfy the `Loose' cut-based identification working point defined in Ref.~\cite{MUON-2018-03}. Signal muons must satisfy the `Medium' identification criteria~\cite{MUON-2018-03} and the PflowTight~\cite{MUON-2018-03} isolation criteria. The longitudinal impact parameter of signal muons is required to be less than 0.5~mm, while the transverse impact parameter significance must be less than 3.  They must also have $\pt>27~\GeV$ and are restricted to the range $|\eta| < 2.5$. The isolation requirement for \enquote{background} muons, used to estimate the background processes with non-prompt muons, is changed to `PflowLoose' and the transverse impact parameter significance is required to be less than 10~\cite{MUON-2018-03}. Background muons are required to fail the signal muon selection to ensure that the samples of signal and background muons are statistically independent.

Jets are reconstructed using the \antikt algorithm~\cite{Cacciari:2008gp, Fastjet}, with a radius parameter of \(R=0.4\), using particle-flow objects~\cite{PERF-2015-09} as input. The jets are calibrated as described in Ref.~\cite{JETM-2018-05} and required to have $\pt > 25~\GeV$ and $|\eta| \le 4.5$. Contamination from jets originating in \pileup collisions is reduced by using the jet-vertex tagger (JVT) algorithm~\cite{PERF-2014-03}.  The JVT discriminant is used to identify jets originating from the hard-scatter vertex through the use of reconstructed parameters of tracks and vertices by requiring  \text{JVT} > 0.50 for jets with $\pt <$ 60~\GeV and $|\eta| <$ 2.5.  In order to suppress contributions from background processes that involve top quarks or leptonic $b$-hadron decays, the DL1r classification algorithm based on recurrent neural networks~\cite{FTAG-2019-07} is used to identify jets originating from  $b$-quarks, referred to as \enquote{\(b\)-jets}.  The \(b\)-jet efficiency working point is  \SI{85}{\percent} in $\ttbar$ events with an expected rejection factor (defined as the inverse of the efficiency) of about 40 for light-flavour jets, and about 2.9 for jets originating from charm quarks for jets with \(\pT > 20~\GeV\) and $|\eta| < 2.5$~\cite{FTAG-2018-01, FTAG-2019-02, FTAG-2020-08}.

The missing transverse momentum, with magnitude $\MET$, is calculated using the negative vector sum of the transverse momenta of all of the selected and calibrated objects in the event. The particle-flow-based algorithm is used, which utilizes  electrons, muons, jets, and \pileup resistant track-based variables for the soft energy term~\cite{ATLAS:2024cmj, ATLAS-CONF-2018-023}. The $ee$ channel utilizes the object-based \met significance ($\mathcal{S}$), calculated by including the expected resolutions for all the objects used in the \met calculation~\cite{ATLAS-CONF-2018-038} and is used to reduce background events with genuine $\MET$.

An object overlap removal procedure is used to avoid labelling the same detector signature as more than one object. The baseline muons and electrons are considered for the overlap removal. First, this procedure removes any electron if it shares an ID track with another higher $\pt$ electron.  Second, electrons sharing their track with a muon candidate are removed. Then, a jet is removed if it overlaps with an electron within $\Delta R_y = \sqrt{(\Delta y)^2+(\Delta \phi)^2} = 0.2$ distance, unless it is a $b$-jet and the electron \pt\ is below 100~\GeV, in which case the electron is removed and any electrons within $\Delta R_y = 0.4$ of a remaining jet are removed. Any jet that is within $\Delta R_y = 0.2$ of a muon and has less than three associated tracks is removed, unless it is a $b$-jet. Finally, remaining muons are removed if their track is within $\Delta R_y = 0.4$ of a remaining jet.

\section{Analysis strategy}
 
The signal region (SR) for both the $ee$ and $e\mu$ channels is defined in order to optimize the sensitivity to the Phenomenological Type-I Seesaw  model with Majorana neutrino masses above 1~\TeV. The contribution of SM same-sign $WW$ and $WZ$ processes is estimated from the MC simulation and normalized to data in dedicated CRs.  The variable that most effectively discriminates between signal and background events is the transverse momentum of the subleading $\pt$ lepton ($\pt^{\ell_2}$), as the signal is expected to populate the high  $\pt^{\ell_2}$ region.
 
Candidate events are required to have a same-sign signal electron pair or electron-muon pair. At least two jets are required and one of the jets is required to have $p_{T}$ larger than 30~\GeV to reduce the contamination from \pileup jets. Events with $b$-jets are discarded. The two  jets with the highest \pt are required to have $m_{jj}>500~\GeV$ and $ \vert \Delta y_{jj} \vert > 2$. Events with electrons in the forward regions of the detector are rejected in the $ee$ channel by requiring $|\eta^{e}| < 2$  to suppress backgrounds with an electron charge mis-identification. The dielectron mass $m_{ee}$ is required to be greater than 20~\GeV in the $ee$ channel.
 
The $ee$ SR is defined by requiring $\mathcal{S}$ to have an upper bound of 4.5 to reduce the same-sign $WW$ background process with neutrinos in the final state. Events with $ee$ invariant mass close to the $Z$ boson mass, $|m_{ee} -m_Z| < 15$~\GeV,  are vetoed in order to reduce the SM Drell--Yan background.   The $e\mu$ SR is defined by requiring a large azimuthal angle separation between the electron and the muon, $\vert \Delta \phi_{e\mu} \vert > 2.0 $. No explicit requirement on $\MET$ is applied in the $e\mu$ channel as no further improvement in signal sensitivity was found for $m_N$ greater than 1~\TeV; due to limited  $\MET$ resolution.  Events with an additional baseline muon or electron are vetoed to reduce the contribution from the $ZZ$ background production.
 
The same-sign $WW$ background normalization is estimated from a CR that has the same requirements as the SR, except for $\mathcal{S}  > 4.5$ and $\vert \Delta \phi_{e\mu} \vert<2.0$ in the $ee$ and $e\mu$ channels, respectively, to ensure no overlap of events with the SR. Additional requirements are imposed to maximize the purity of this background in the CR by increasing the minimum transverse momentum requirements on the leading ($\pt^{j_1}$) and subleading ($\pt^{j_2}$) jets to 45 and 30~\GeV, respectively. The leading electron transverse momentum  ($\pt^{\ell_1}$) is required to be less than 250~\GeV in the $ee$ channel in order to reduce the signal contamination in this CR.

\begin{table}[t!bp]
\centering
\caption{Summary of the event selection requirements in the SR and CRs in the $ee$ and $e\mu$ channels. $N_{\ell}$ denotes the number of selected leptons.}
\label{tab:preselection}
\begin{tabular}{c|c|c|c|c}
\hline\hline
Channel & Variable               & SR            & $W^{\pm}W^{\pm}$ CR           & $WZ$ CR \\ \hline
\multirow{4}{*}{$ee$/$e\mu$}  &  $N_{\ell}$        & \multicolumn{2}{c|}{=2}  & =3               \\
&   $|\Delta y_{jj}|$        & \multicolumn{3}{c}{> 2}                     \\
&   $m_{jj}$           & \multicolumn{3}{c}{> 500 \GeV}                   \\
&   $m_{\ell\ell\ell}$  & --             & --                & > 106 \GeV \\
\hline
\multirow{7}{*}{$ee$} &   $|m_{\ell\ell}-m_Z|$  & \multicolumn{2}{c|}{> 15 \GeV}        & --    \\
&       $|\eta_{\ell}|$          & \multicolumn{3}{c}{<2}                     \\
&  $m_{\ell\ell}$   & \multicolumn{3}{c}{> 20 \GeV}                    \\
& $\pt^{\ell_1}$    &   --           & < 250              & --    \\
& $\pt^{j_1}$     &  > 30 \GeV           & > 45 \GeV               & > 30 \GeV  \\
& $\pt^{j_2}$     &  > 25 \GeV            & > 30 \GeV              & > 25 \GeV   \\
&  $\mathcal{S}$                  & < 4.5           & > 4.5              & --    \\
\hline
\multirow{3}{*}{$e\mu$}  &    $\pt^{j_1}$      &  > 30 \GeV           & > 45 \GeV               & > 45 \GeV  \\
&  $\pt^{j_2}$     &  > 25 \GeV            & > 30 \GeV              & > 30 \GeV   \\
&  $\vert \Delta \phi_{e\mu} \vert $               & > 2.0           & < 2.0              & --    \\
\hline\hline
\end{tabular}
\end{table}
 
Processes in which $WZ$ bosons are produced in association with jets are another large source of background events. The $WZ$ CR is defined by selecting events with three leptons, two of which have opposite charge in order to be compatible with a $Z$ boson decay. The third lepton is required to satisfy $\pt > 15~\GeV$. Events containing a fourth electron or muon are removed to reject events from the $ZZ$ background process. The invariant mass of the three leptons, $m_{\ell\ell\ell}$, is required to be greater than 106~\GeV in order to reduce the fraction of events with non-prompt leptons originating from the $Z$+jets background. The event selection requirements used to define the SR and the two CRs in the $ee$ and $e\mu$ channels are summarized in Table~\ref{tab:preselection}.

The remaining sources of background are estimated using data-driven methods or from MC simulation. The non-prompt background contribution is evaluated using the so-called fake-factor method~\cite{EGAM-2019-01}. Fake factors are measured as a function of electron or muon $\pt$ and $\eta$ in a dedicated region enriched in non-prompt leptons that selects collision-data events containing jets recoiling against the non-prompt lepton candidate. The jet-lepton back-to-back topology is selected by requiring the difference in azimuthal angle between the lepton and the jet, $\vert \Delta\phi_{\ell{j}} \vert$, to be larger than 2.8.  In this region, the contribution of events containing electrons or muons from $W$ or $Z$  boson decays is reduced with requirements on kinematic variables as described in Ref.~\cite{EGAM-2019-01}. The remaining events containing electrons or muons from $W$, $Z$, or top-quark decays, as well as from photon conversion in $\gamma$+jet processes are subtracted from this region using MC simulation. The fake factors for electrons (muons) are measured independently as the ratio between the number of signal electrons (muons) and background electrons (muons).  Finally, the non-prompt background contribution in the SR is estimated by re-scaling data from the same SR event selection with background leptons by the corresponding fake factors~\cite{EGAM-2019-01}.

Another significant source of background is from opposite-sign lepton pair SM events in which one of the two lepton charges has been misidentified.  This process occurs more frequently for electrons mainly due to the presence of bremsstrahlung radiation followed by electron-positron pair production. This background, referred to as \textit{charge flip}, is estimated by re-weighting opposite-sign events in data by a factor representing the probability for charge flip to occur to one of the electrons~\cite{EGAM-2018-01}. The charge flip rate is derived as a function of electron $\pt$ and $\eta$, using simulated $Z\rightarrow ee$ events which are corrected to the data observations through the application of dedicated scale-factors. Contributions from low-energy multijet events are removed by requiring the electron pairs to satisfy $m_{\ell\ell} >$ 20~\GeV. Contributions from electrons originating from final state radiation are subtracted using MC simulation. The charge flip rates are then measured as the fraction of the number of electrons with wrongly--assigned charge over all electrons in this region.
 
The background contribution from $\ell\gamma jj$ events, where the prompt photon $\gamma$ is misreconstructed as an electron, is simulated using the $V\gamma$ MC simulation.  Other minor backgrounds, including different combinations of bosons, tops, jets and photons, are also estimated from MC simulation.
 
\section{Results}
 
\begin{figure}[t!bp]
\centering
\subfigure[]{\includegraphics[width=0.32\textwidth]{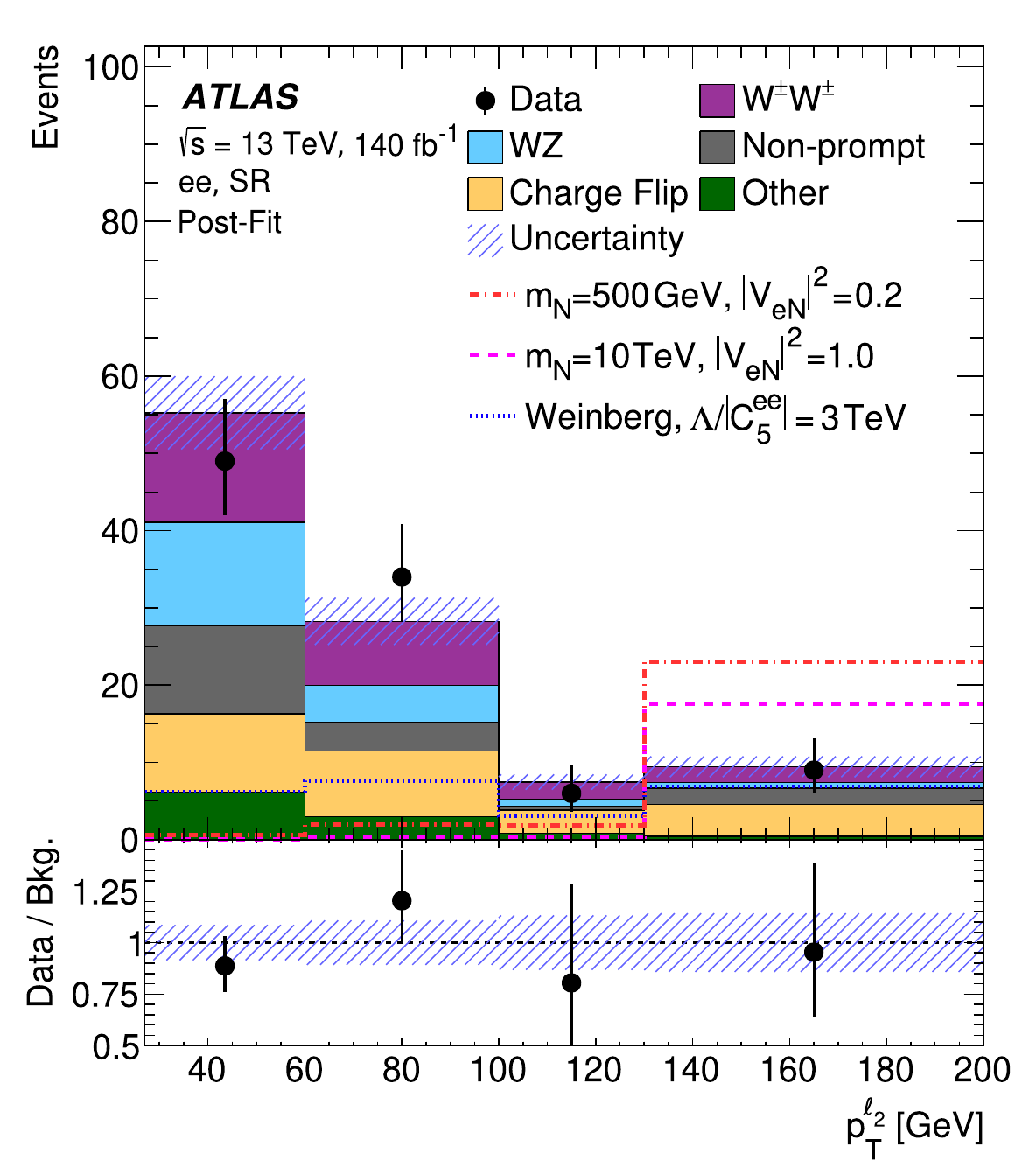}}
\subfigure[]{\includegraphics[width=0.32\textwidth]{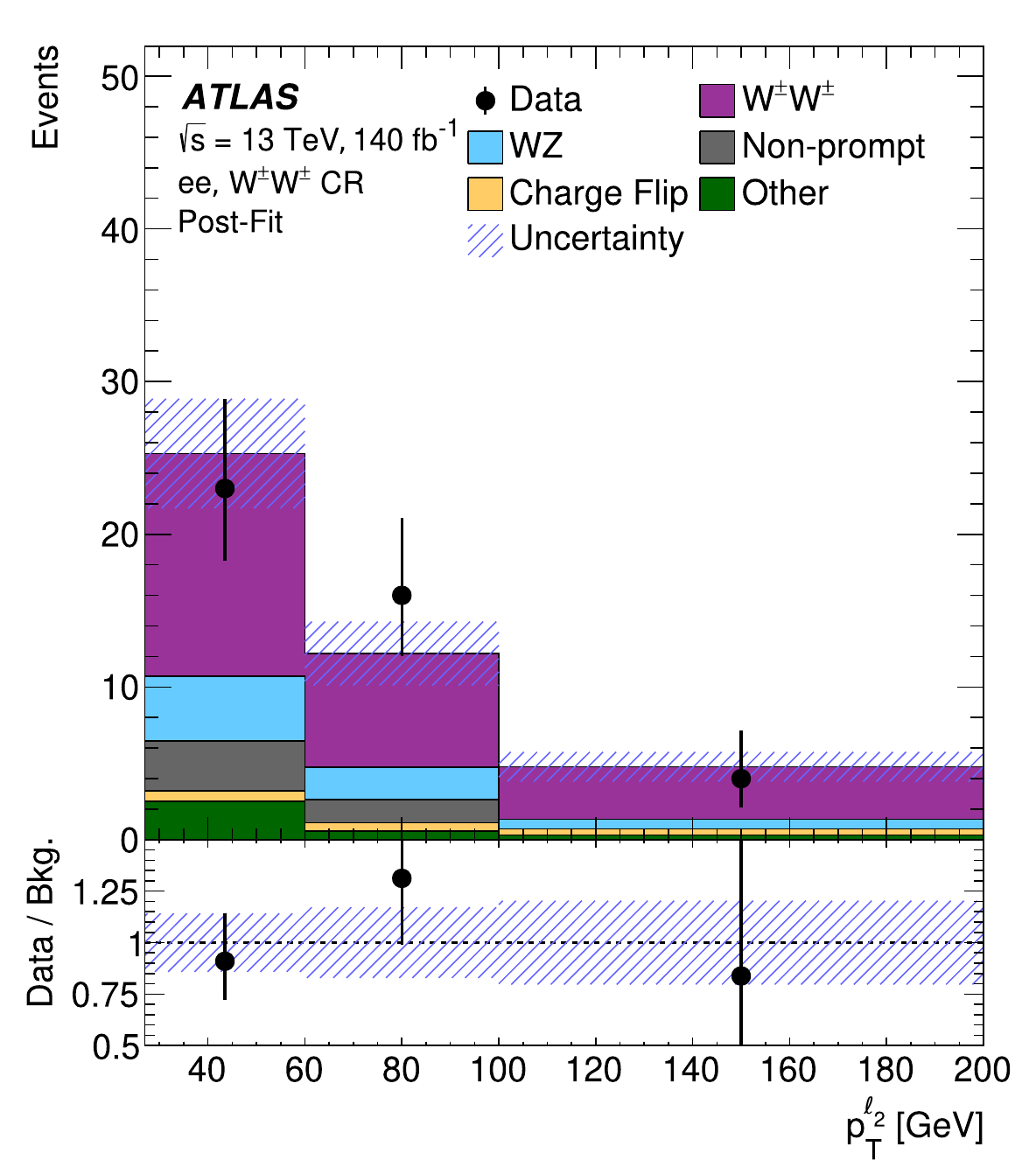}}
\subfigure[]{\includegraphics[width=0.32\textwidth]{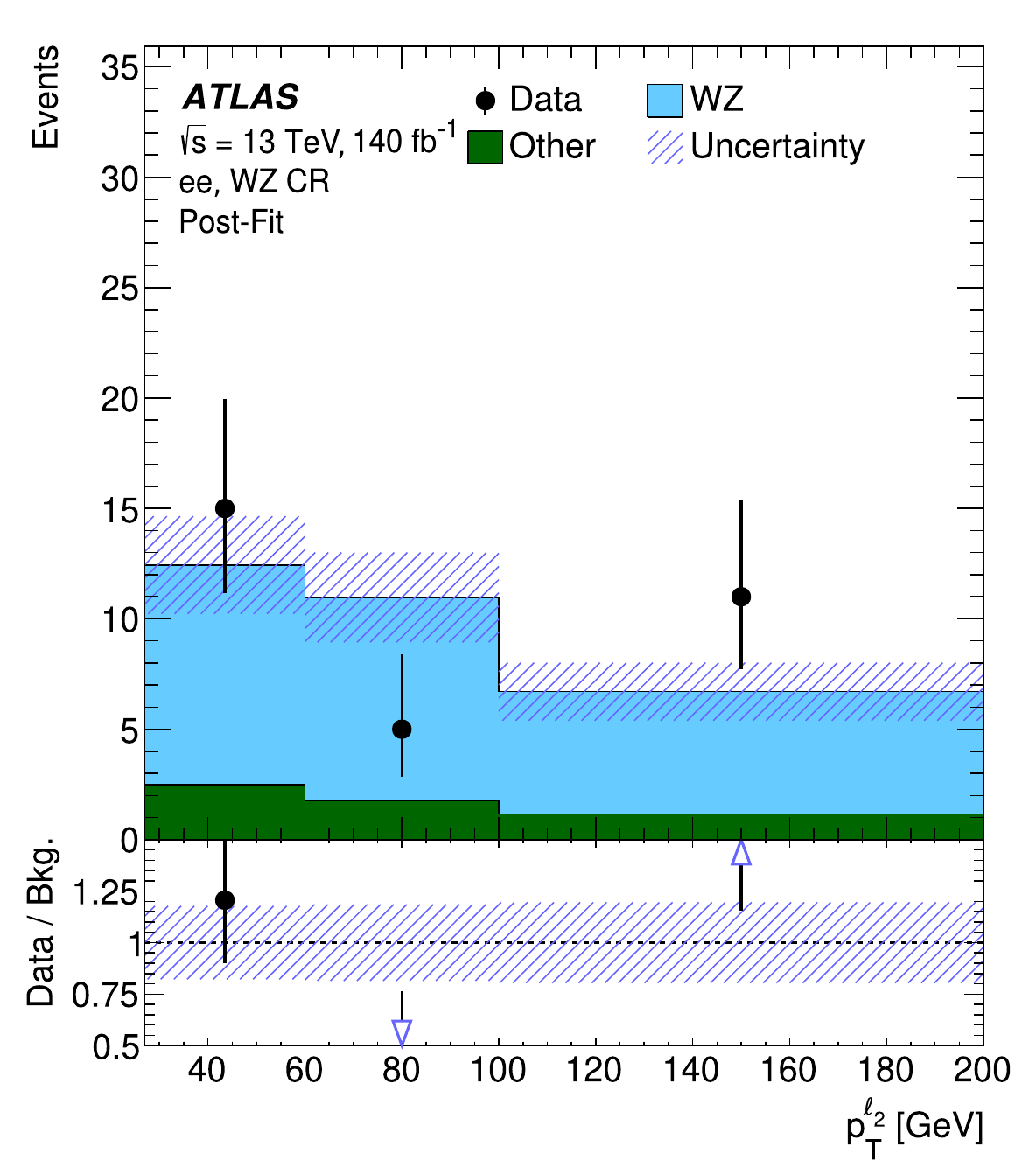}}
\subfigure[]{\includegraphics[width=0.32\textwidth]{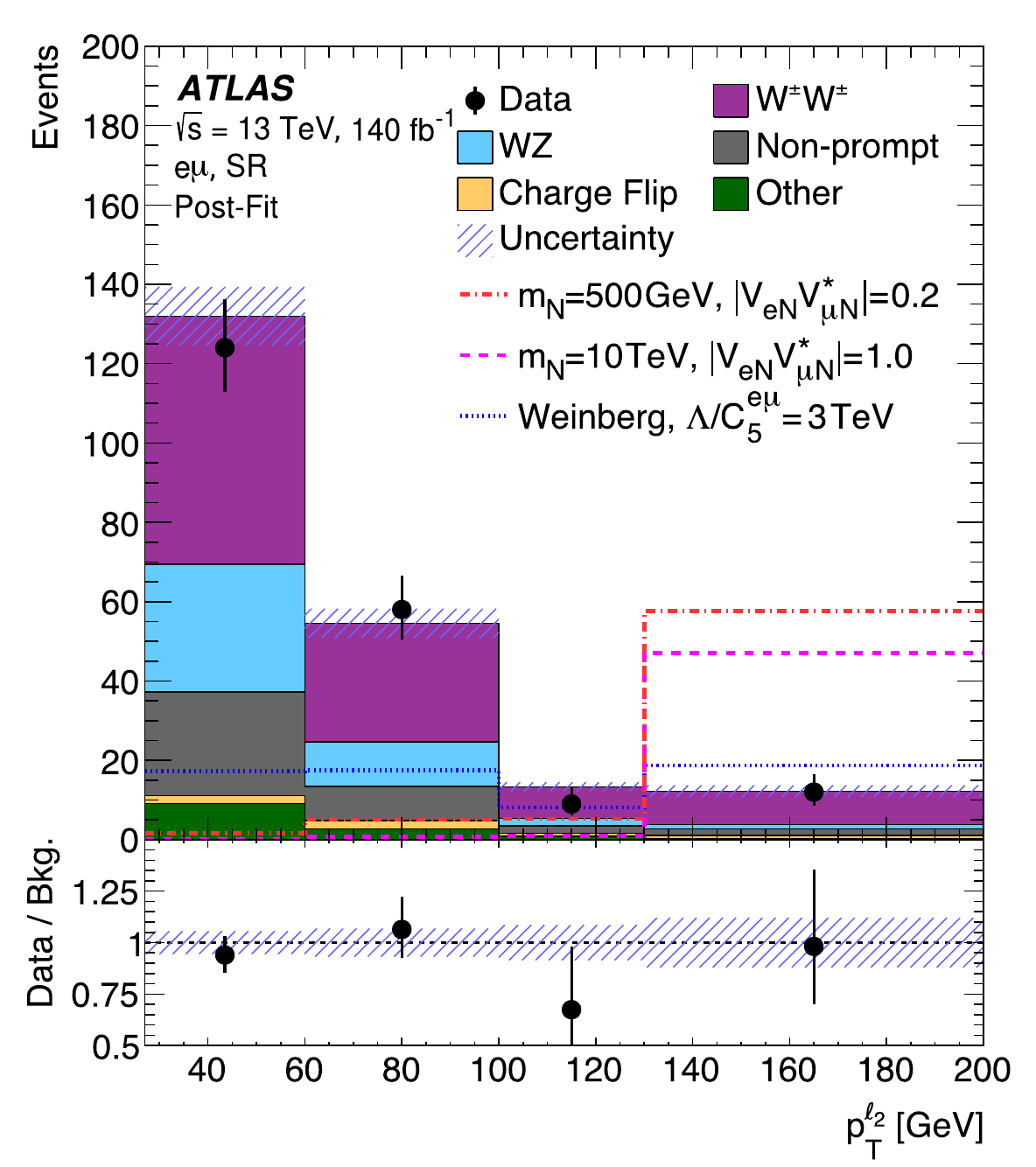}}
\subfigure[]{\includegraphics[width=0.32\textwidth]{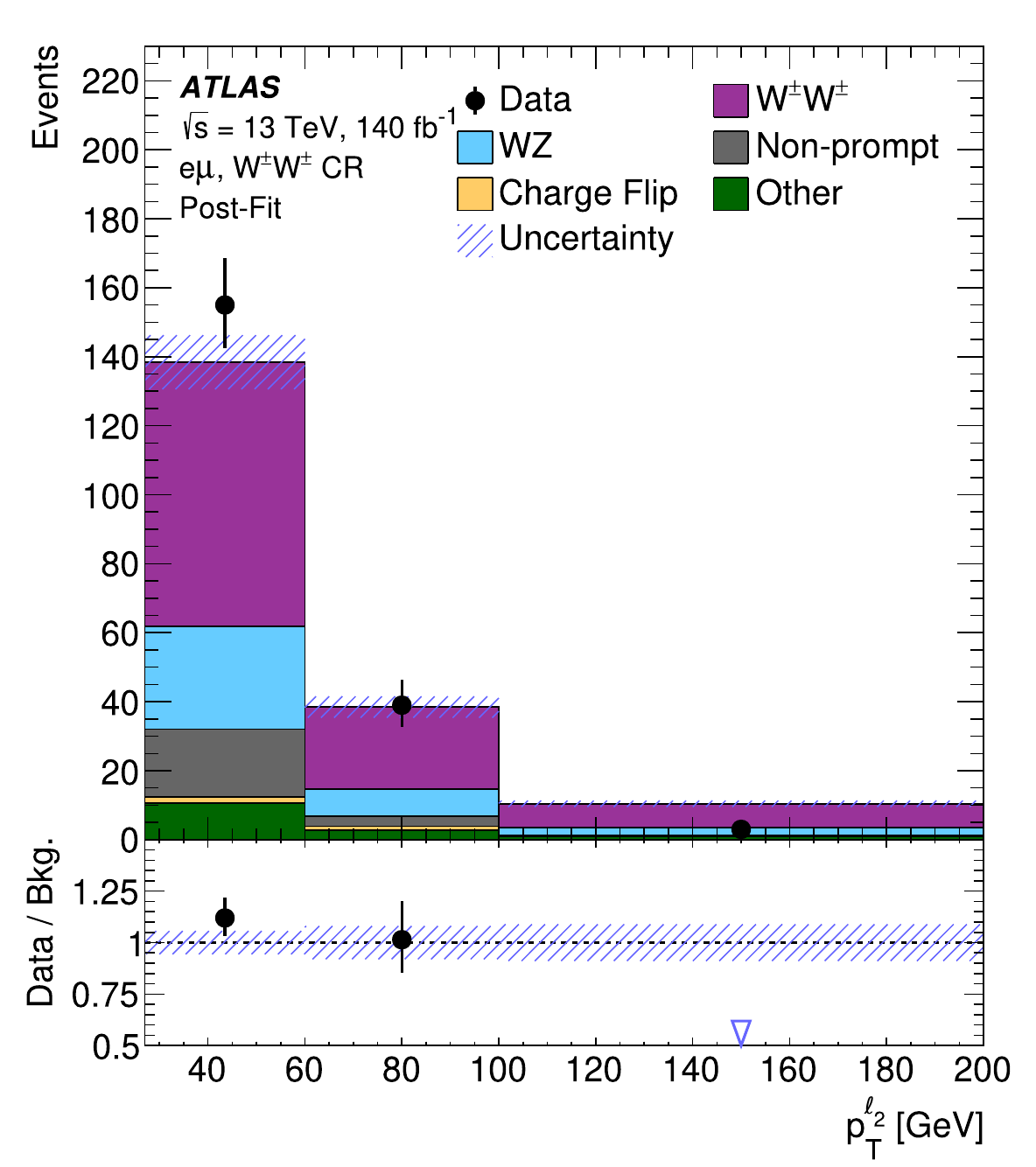}}
\subfigure[]{\includegraphics[width=0.32\textwidth]{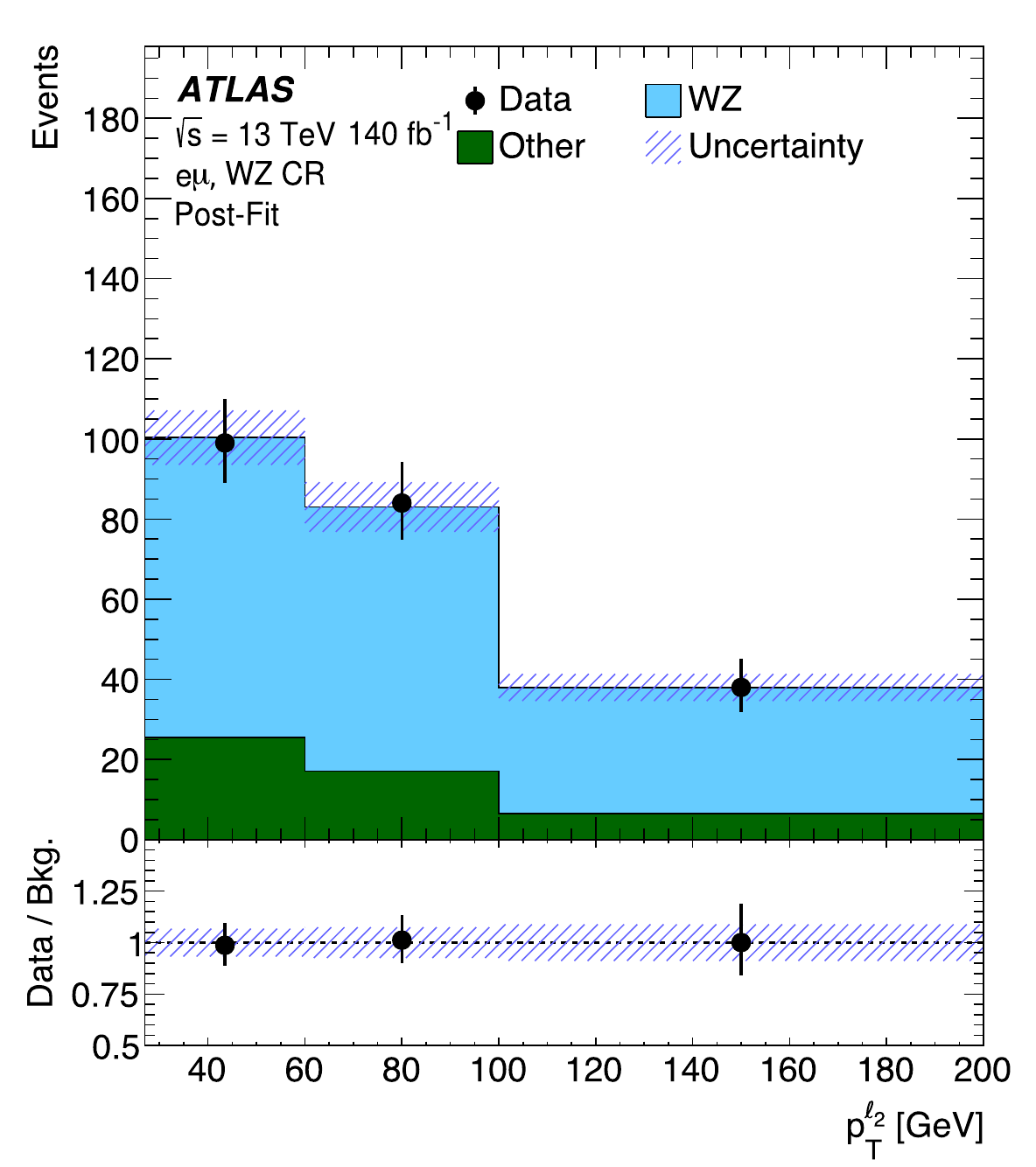}}
\caption{Distributions of the $\pt^{\ell_2}$ in (a) the $ee$ SR, (b) the $ee$ same-sign $WW$ CR,  (c) the $ee$ $WZ$ CR, (d) the $e\mu$ SR, (e) the $e\mu$ same-sign $WW$ CR,  and  (f) the $e\mu$ $WZ$ CR. The predicted yields are shown with their best fit normalization and shape from the fit for the SM background-only hypothesis. The shaded area surrounding the background expectation represents the total uncertainties in the predicted yields. The ratios of the observed yields to the total SM predictions are shown by the points in the bottom panels.  `Other' includes contributions  mainly from the $\ell\gamma jj$, $ZZ$ and \tZq background processes.  Benchmark signal predictions are shown as hatched lines. The last bin includes the overflow.}
\label{fig:emu_plots}
\end{figure}

A binned maximum-likelihood fit is performed separately in the $ee$ and $e\mu$ channels using the $\pt^{\ell_2}$ distributions, as shown in Figure~\ref{fig:emu_plots}. The SR and the two CRs are fitted simultaneously for each channel. The signal, same-sign $WW$ and $WZ$ background normalizations are kept as unconstrained parameters in the fits.  The systematic uncertainties are included as nuisance parameters~\cite{Rolke:2004mj}, with Gaussian priors correlated across the SR and CRs. The nuisance parameters are profiled in the fit with the shape and normalization of each distribution varying within the specified constraints. The background-dominated CRs reduce the impact of the systematic uncertainties after the fit.
 
The results are driven by the statistical uncertainty of the data in the tail of the $\pt^{\ell_2}$ distribution in the SR and none of the considered systematic uncertainties have significant impact on the sensitivity of this search. The expected limits improve by about 5\% and 3\% in the $ee$ and $e\mu$ channels, respectively, if the systematic uncertainties are not included.  The largest systematic uncertainties considered are briefly discussed in the following.
 
The theoretical uncertainties in the physics modelling of the signal and SM same-sign $WW$ and $WZ$ background processes are estimated by varying the factorization ($\mu_F$) and renormalization ($\mu_R$) scales, the strong coupling constant $\alpha_S$ and the choice of the PDF. The $\mu_F$ and $\mu_R$ are varied independently by factors of 0.5 and 2 with respect to their nominal values. The envelope of the resulting  variations, excluding the two variations where one scale is varied up and the other one down,  is used as a single uncertainty.  The PDF uncertainty is obtained from the standard variations of the \NNPDF PDF set and by varying the nominal $\alphas=0.118$ within its uncertainty of $\pm 0.001$. The impact of the NLO QCD corrections for the EW same-sign $WW$ production was assessed by comparing the nominal LO sample with an alternative NLO sample generated with \SHERPA[2.2.11].
 
The uncertainties in the non-prompt electron and muon background processes are studied in detail. Three sources of systematic uncertainty are considered including the statistical error on the fake factors, uncertainties on the composition of the fake factor CR obtained by varying $b$-jet requirement in this region, and uncertainties related to the prompt lepton contribution in the fake factor CR. Uncertainties in the data-driven charge flip background are described in Ref.~\cite{EGAM-2018-01}. Uncertainty sources related to electron and muon reconstruction and energy calibration, jet energy calibration, $b$-jet identification, rejection of \pileup jets, and missing transverse momentum reconstruction were considered as well.

The distributions of $\pt^{\ell_2}$ in the SR, same-sign $WW$ CR, and $WZ$ CR in the $ee$ and $e\mu$ channels are shown in Figure~\ref{fig:emu_plots}. The $ee$ channel background normalization factors for the same-sign $WW$ and $WZ$ background processes are 1.13$\pm$0.30 and 0.82$\pm$0.21, respectively. The same-sign $WW$ and $WZ$ normalization factors in the $e\mu$ channel are 1.20$\pm$0.15 and 0.65$\pm$0.12, respectively.  The obtained normalization factors are consistent with the values reported in Refs.~\cite{ATLAS:2023sua,STDM-2017-23}. The post-fit SM background predictions under the background-only hypothesis, together with the data yields in the SRs of the $ee$ and $e\mu$ channels are shown in Table~\ref{tab:yields}.

\begin{table}[tbp]
\centering
\caption{Post-fit SM background yields under the background-only hypothesis and observed data events in the SR in the $ee$ and $e\mu$ channels. The total uncertainties in the predicted yields are shown. `Other' includes contributions mainly from the $\ell\gamma jj$, $ZZ$ and \tZq background processes. \label{tab:yields}}
\begin{tabular}{c|r@{$\,\pm\,$}l|r@{$\,\pm\,$}l}
\hline
\hline
Channel & \multicolumn{2}{c|}{$ee$}  & \multicolumn{2}{c}{$e \mu$}  \\
\hline
Same-sign $WW$ &  26.8 & 7.3 & 109 & 13 \\
WZ &  20.0 & 5.6 & 46 & 11 \\
Non-prompt   & 17.7 & 2.8 & 38.4 & 5.0 \\
Charge flip & 25.8 & 7.4 & 5.48 & 0.82 \\
Other   & 10.3 & 5.3 & 13.1 & 6.4 \\
\hline
Total SM  & 101 & 8 & 212 & 11 \\
\hline
Data   & \multicolumn{2}{c|}{98} & \multicolumn{2}{c}{203} \\
\hline
\end{tabular}
\end{table}

\begin{figure}[t!bp]
\centering
\subfigure[]{\includegraphics[width=.45\textwidth]{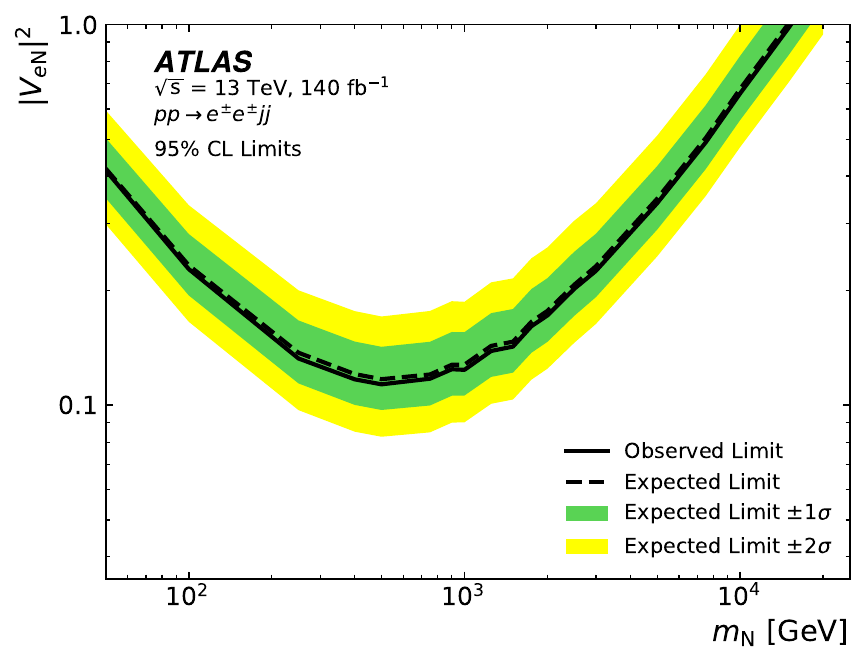}}
\subfigure[]{\includegraphics[width=.45\textwidth]{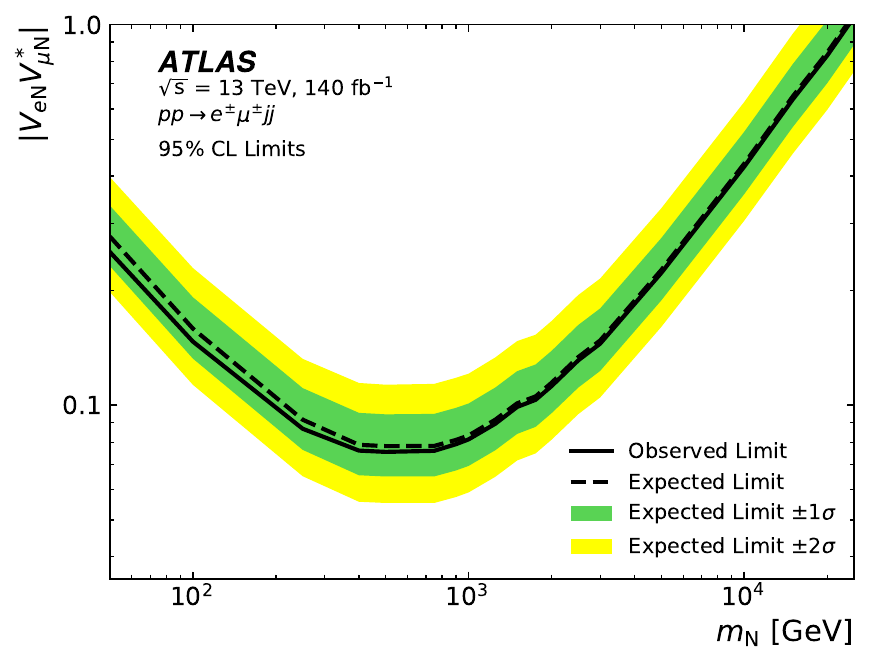}}
\ \caption{Observed and expected 95\% CL upper limits on the heavy Majorana neutrino mixing element (a) \VeN and (b) \VemuN as a function of $m_N$ in the Phenomenological Type-I Seesaw model. The one and two standard deviation bands of the expected limit are indicated in green and yellow, respectively. }
\label{fig:limitplot}
\end{figure}

No significant excess of events above the SM background expectation is observed. The 95\% confidence level (CL) upper limits on \VeN and $\vert m_{ee} \vert$, and on \VemuN and $\vert m_{e\mu} \vert$ are derived separately in the $ee$ and $e\mu$ channels, respectively, using the $\text{CL}_{\textrm{s}}$ method~\cite{Junk_1999,Read:2002hq}. The asymptotic approximation~\cite{Cowan:2010js}, whose validity was confirmed through studies with pseudo-experiments, is used to derive the upper limits. The observed and expected 95\% CL limits on \VeN and \VemuN in the $ee$ and $e\mu$ channels, respectively, as a function of $m_N$ are shown in Figure~\ref{fig:limitplot}. The small deficit of data with respect to the SM background prediction in the last bin of the $\pt^{\ell_2}$  distribution in the SR of the $e\mu$ channel results in a stronger observed limit by about one standard deviation with respect to the expected limit across the entire range of $m_N$.  The strongest exclusion limits are achieved at $m_N=500$~\GeV, with observed (expected) upper limits of 0.11~(0.12) and 0.075~(0.078) on \VeN  and \VemuN, respectively. The observed (expected) 95\% CL lower limits on $\Lambda/ |C_5^{ee}|$ and $\Lambda/ |C_5^{e\mu}|$ obtained using the dimension-5 Weinberg operator signal are  2.5~\TeV~(2.6~\TeV) and 4.9~\TeV~(4.3~\TeV) in the $ee$ and $e\mu$ channels, respectively. This translates into observed (expected) 95\% CL upper limits of 24~\GeV~(24~\GeV) and 12~\GeV~(14~\GeV) on $\vert m_{ee} \vert$  and  $\vert m_{e\mu} \vert$  in the $ee$ and $e\mu$ channels, respectively.

\begin{figure}[htbp]
\centering
\subfigure[]{\includegraphics[width=.45\textwidth]{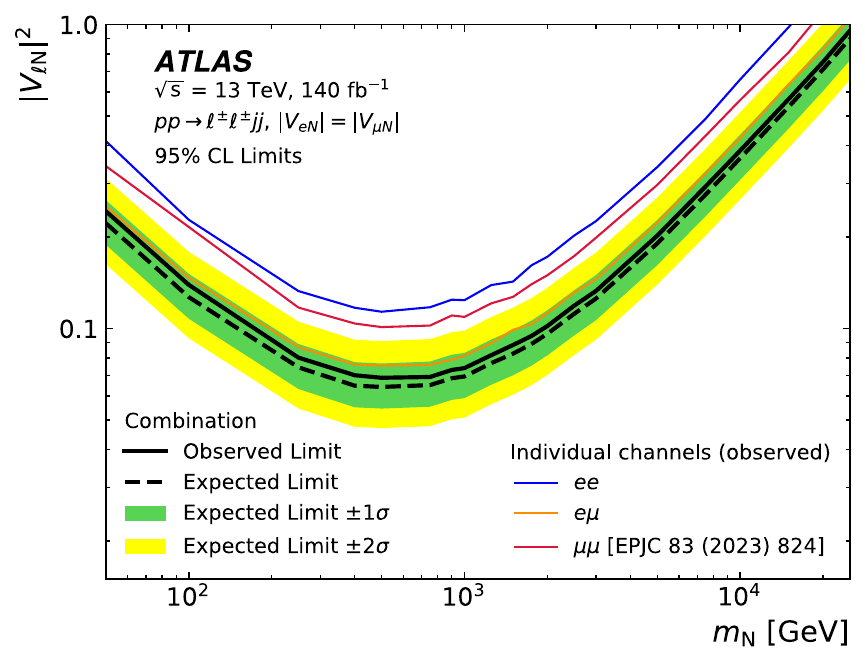}}
\subfigure[]{\includegraphics[width=.37\textwidth]{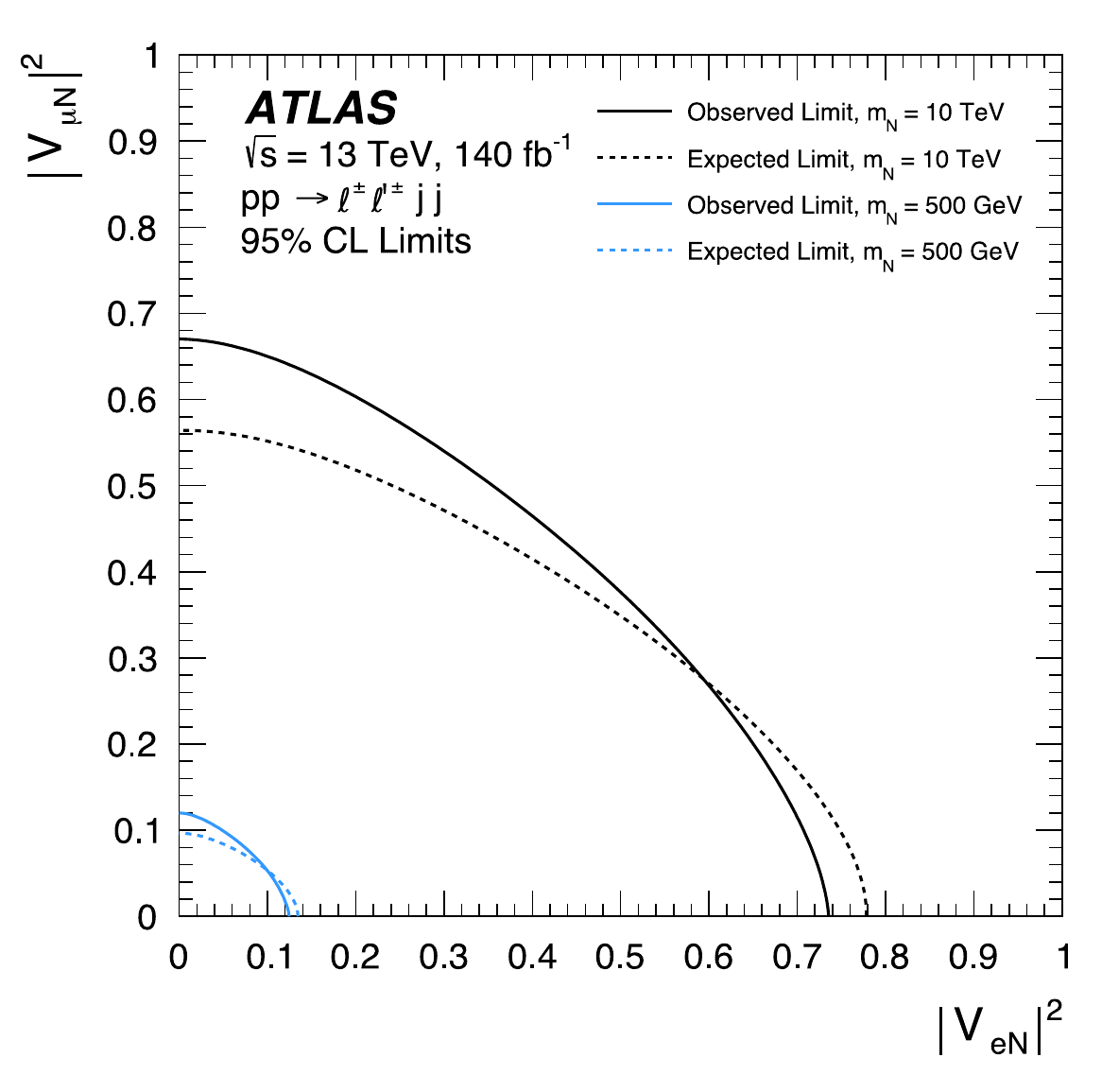}}
\ \caption{(a) Observed and expected 95\% CL upper limits on the heavy Majorana neutrino mixing element \VlN as a function of $m_N$.  The limits on the \VlN are obtained assuming $\VmuN = \VeN$. The observed limits on the \VlN obtained separately in the $ee$, $e\mu$, and $\mu\mu$~\cite{EXOT-2020-06} channels are also shown for comparison. The one and two standard deviation bands of the expected limit are indicated in green and yellow, respectively. (b) Two-dimensional observed and expected 95\% CL intervals on \VeN and \VmuN for $m_N=500~\GeV$ and $10$~\TeV in the Phenomenological Type-I Seesaw model.  }
\label{fig:limitplot2}
\end{figure}

The first statistical combination of searches for heavy Majorana neutrinos in the $ee$, $e\mu$, and $\mu\mu$~\cite{EXOT-2020-06} channels is also reported. The SRs and CRs are defined to be non-overlapping  across the channels. The normalization parameters of the same-sign $WW$ and $WZ$ background processes are floated separately in each channel in the simultaneous fit. The correlations between the systematic uncertainties across the channels are taken into account.  Figure~\ref{fig:limitplot2}(a) shows the observed and expected 95\% CL upper limits on the heavy Majorana neutrino mixing element \VlN assuming $\VmuN = \VeN$ in the Phenomenological Type-I Seesaw model. The observed limits on the \VlN obtained separately in the $ee$, $e\mu$, and $\mu\mu$ channels are also shown. The combined observed (expected) limit is about 32\%  (24\%) more stringent compared to the respective limits obtained in the $\mu\mu$~\cite{EXOT-2020-06} channel. The results are also interpreted by relaxing the $\VmuN = \VeN$  assumption. The two-dimensional observed and expected 95\% CL intervals on the \VeN and \VmuN parameters are shown in Figure~\ref{fig:limitplot2}(b) for $m_N=500~\GeV$ and $10$~\TeV.

\section{Conclusion}
 
The first searches for heavy Majorana neutrinos in scattering of same-sign $W$ boson pairs in the $ee$ and $e\mu$ channels are reported. The dataset corresponds to proton--proton collisions at  $\sqrt{s}= 13~\TeV$ at the LHC with an integrated luminosity of 140~fb$^{-1}$, collected with the ATLAS detector during 2015--2018.  No significant excess of events with respect to the SM background prediction is observed. The results are interpreted in a benchmark scenario of the Phenomenological Type-I Seesaw model.  Bounds are reported on  $\vert V_{e N} \vert^{2}$  and $\vert V_{e N} V^{*}_{\mu N} \vert$  in the heavy Majorana neutrino mass range between 50~\GeV and 25~\TeV, where $V_{\ell N}$ is the matrix element describing the mixing of the heavy Majorana neutrino mass eigenstate with the Standard Model of flavour $\ell$, with $\ell = e$ or  $\mu$. The first statistical combination of the $ee$,  $e\mu$, and $\mu\mu$ channels is performed. The combined observed (expected) limit on  \VlN, assuming $\VmuN = \VeN$,  is about 32\%  (24\%) more stringent compared to the respective limits obtained in the $\mu\mu$ channel. The sensitivity to the Weinberg operator is investigated and constraints on the effective $ee$ and $e\mu$ Majorana neutrino masses are reported. The observed (expected) 95\% CL upper limits on $\vert m_{ee} \vert$  and  $\vert m_{e\mu} \vert$ obtained using the dimension-5 Weinberg operator signal are 24~\GeV~(24~\GeV) and 12~\GeV~(14~\GeV) in the $ee$ and $e\mu$ channels, respectively.

\section*{Acknowledgements}


We thank CERN for the very successful operation of the LHC and its injectors, as well as the support staff at
CERN and at our institutions worldwide without whom ATLAS could not be operated efficiently.
 
The crucial computing support from all WLCG partners is acknowledged gratefully, in particular from CERN, the ATLAS Tier-1 facilities at TRIUMF/SFU (Canada), NDGF (Denmark, Norway, Sweden), CC-IN2P3 (France), KIT/GridKA (Germany), INFN-CNAF (Italy), NL-T1 (Netherlands), PIC (Spain), RAL (UK) and BNL (USA), the Tier-2 facilities worldwide and large non-WLCG resource providers. Major contributors of computing resources are listed in Ref.~\cite{ATL-SOFT-PUB-2023-001}.
 
We gratefully acknowledge the support of ANPCyT, Argentina; YerPhI, Armenia; ARC, Australia; BMWFW and FWF, Austria; ANAS, Azerbaijan; CNPq and FAPESP, Brazil; NSERC, NRC and CFI, Canada; CERN; ANID, Chile; CAS, MOST and NSFC, China; Minciencias, Colombia; MEYS CR, Czech Republic; DNRF and DNSRC, Denmark; IN2P3-CNRS and CEA-DRF/IRFU, France; SRNSFG, Georgia; BMBF, HGF and MPG, Germany; GSRI, Greece; RGC and Hong Kong SAR, China; ISF and Benoziyo Center, Israel; INFN, Italy; MEXT and JSPS, Japan; CNRST, Morocco; NWO, Netherlands; RCN, Norway; MNiSW, Poland; FCT, Portugal; MNE/IFA, Romania; MESTD, Serbia; MSSR, Slovakia; ARRS and MIZ\v{S}, Slovenia; DSI/NRF, South Africa; MICINN, Spain; SRC and Wallenberg Foundation, Sweden; SERI, SNSF and Cantons of Bern and Geneva, Switzerland; NSTC, Taipei; TENMAK, T\"urkiye; STFC, United Kingdom; DOE and NSF, United States of America.
 
Individual groups and members have received support from BCKDF, CANARIE, CRC and DRAC, Canada; CERN-CZ, PRIMUS 21/SCI/017, FORTE and PRIMUS, Czech Republic; COST, ERC, ERDF, Horizon 2020, ICSC-NextGenerationEU and Marie Sk{\l}odowska-Curie Actions, European Union; Investissements d'Avenir Labex, Investissements d'Avenir Idex and ANR, France; DFG and AvH Foundation, Germany; Herakleitos, Thales and Aristeia programmes co-financed by EU-ESF and the Greek NSRF, Greece; BSF-NSF and MINERVA, Israel; Norwegian Financial Mechanism 2014-2021, Norway; NCN and NAWA, Poland; La Caixa Banking Foundation, CERCA Programme Generalitat de Catalunya and PROMETEO and GenT Programmes Generalitat Valenciana, Spain; G\"{o}ran Gustafssons Stiftelse, Sweden; The Royal Society and Leverhulme Trust, United Kingdom.
 
In addition, individual members wish to acknowledge support from CERN: European Organization for Nuclear Research (CERN PJAS); Chile: Agencia Nacional de Investigaci\'on y Desarrollo (FONDECYT 1190886, FONDECYT 1210400, FONDECYT 1230812, FONDECYT 1230987); China: Chinese Ministry of Science and Technology (MOST-2023YFA1605700), National Natural Science Foundation of China (NSFC - 12175119, NSFC 12275265, NSFC-12075060); Czech Republic: Czech Science Foundation (GACR - 24-11373S), Ministry of Education Youth and Sports (FORTE CZ.02.01.01/00/22\_008/0004632), PRIMUS Research Programme (PRIMUS/21/SCI/017); EU: H2020 European Research Council (ERC - 101002463); European Union: European Research Council (ERC - 948254, ERC 101089007), Horizon 2020 Framework Programme (MUCCA - CHIST-ERA-19-XAI-00), European Union, Future Artificial Intelligence Research (FAIR-NextGenerationEU PE00000013), Italian Center for High Performance Computing, Big Data and Quantum Computing (ICSC, NextGenerationEU); France: Agence Nationale de la Recherche (ANR-20-CE31-0013, ANR-21-CE31-0013, ANR-21-CE31-0022, ANR-22-EDIR-0002), Investissements d'Avenir Labex (ANR-11-LABX-0012); Germany: Baden-Württemberg Stiftung (BW Stiftung-Postdoc Eliteprogramme), Deutsche Forschungsgemeinschaft (DFG - 469666862, DFG - CR 312/5-2); Italy: Istituto Nazionale di Fisica Nucleare (ICSC, NextGenerationEU), Ministero dell'Università e della Ricerca (PRIN - 20223N7F8K - PNRR M4.C2.1.1); Japan: Japan Society for the Promotion of Science (JSPS KAKENHI JP21H05085, JSPS KAKENHI JP22H01227, JSPS KAKENHI JP22H04944, JSPS KAKENHI JP22KK0227); Netherlands: Netherlands Organisation for Scientific Research (NWO Veni 2020 - VI.Veni.202.179); Norway: Research Council of Norway (RCN-314472); Poland: Polish National Agency for Academic Exchange (PPN/PPO/2020/1/00002/U/00001), Polish National Science Centre (NCN 2021/42/E/ST2/00350, NCN OPUS nr 2022/47/B/ST2/03059, NCN UMO-2019/34/E/ST2/00393, UMO-2020/37/B/ST2/01043, UMO-2021/40/C/ST2/00187, UMO-2022/47/O/ST2/00148, UMO-2023/49/B/ST2/04085); Slovenia: Slovenian Research Agency (ARIS grant J1-3010); Spain: Generalitat Valenciana (Artemisa, FEDER, IDIFEDER/2018/048), Ministry of Science and Innovation (MCIN \& NextGenEU PCI2022-135018-2, MICIN \& FEDER PID2021-125273NB, RYC2019-028510-I, RYC2020-030254-I, RYC2021-031273-I, RYC2022-038164-I), PROMETEO and GenT Programmes Generalitat Valenciana (CIDEGENT/2019/023, CIDEGENT/2019/027); Sweden: Swedish Research Council (Swedish Research Council 2023-04654, VR 2018-00482, VR 2022-03845, VR 2022-04683, VR 2023-03403, VR grant 2021-03651), Knut and Alice Wallenberg Foundation (KAW 2018.0157, KAW 2018.0458, KAW 2019.0447, KAW 2022.0358); Switzerland: Swiss National Science Foundation (SNSF - PCEFP2\_194658); United Kingdom: Leverhulme Trust (Leverhulme Trust RPG-2020-004), Royal Society (NIF-R1-231091); United States of America: U.S. Department of Energy (ECA DE-AC02-76SF00515), Neubauer Family Foundation.


\printbibliography
 
\clearpage
\input{atlas_authlist}

\end{document}

%% file: atlas_authlist.tex
 
\begin{flushleft}
\hypersetup{urlcolor=black}
{\Large The ATLAS Collaboration}

\bigskip

\AtlasOrcid[0000-0002-6665-4934]{G.~Aad}$^\textrm{\scriptsize 104}$,
\AtlasOrcid[0000-0001-7616-1554]{E.~Aakvaag}$^\textrm{\scriptsize 17}$,
\AtlasOrcid[0000-0002-5888-2734]{B.~Abbott}$^\textrm{\scriptsize 123}$,
\AtlasOrcid[0000-0002-0287-5869]{S.~Abdelhameed}$^\textrm{\scriptsize 119a}$,
\AtlasOrcid[0000-0002-1002-1652]{K.~Abeling}$^\textrm{\scriptsize 56}$,
\AtlasOrcid[0000-0001-5763-2760]{N.J.~Abicht}$^\textrm{\scriptsize 50}$,
\AtlasOrcid[0000-0002-8496-9294]{S.H.~Abidi}$^\textrm{\scriptsize 30}$,
\AtlasOrcid[0009-0003-6578-220X]{M.~Aboelela}$^\textrm{\scriptsize 45}$,
\AtlasOrcid[0000-0002-9987-2292]{A.~Aboulhorma}$^\textrm{\scriptsize 36e}$,
\AtlasOrcid[0000-0001-5329-6640]{H.~Abramowicz}$^\textrm{\scriptsize 154}$
% ...
\bigskip
\\

% ...
$^{17}$Department for Physics and Technology, University of Bergen, Bergen; Norway.\\
% ...
$^{30}$Physics Department, Brookhaven National Laboratory, Upton NY; United States of America.\\
% ...
$^{36}$$^{(a)}$Facult\'e des Sciences Ain Chock, Universit\'e Hassan II de Casablanca;$^{(b)}$Facult\'{e} des Sciences, Universit\'{e} Ibn-Tofail, K\'{e}nitra;$^{(c)}$Facult\'e des Sciences Semlalia, Universit\'e Cadi Ayyad, LPHEA-Marrakech;$^{(d)}$LPMR, Facult\'e des Sciences, Universit\'e Mohamed Premier, Oujda;$^{(e)}$Facult\'e des sciences, Universit\'e Mohammed V, Rabat;$^{(f)}$Institute of Applied Physics, Mohammed VI Polytechnic University, Ben Guerir; Morocco.\\
% ...
$^{45}$Physics Department, Southern Methodist University, Dallas TX; United States of America.\\
% ...
$^{50}$Fakult\"{a}t Physik , Technische Universit{\"a}t Dortmund, Dortmund; Germany.\\
% ...
$^{56}$II. Physikalisches Institut, Georg-August-Universit\"{a}t G\"ottingen, G\"ottingen; Germany.\\
% ...
$^{104}$CPPM, Aix-Marseille Universit\'e, CNRS/IN2P3, Marseille; France.\\
% ...
$^{119}$$^{(a)}$New York University Abu Dhabi, Abu Dhabi;$^{(b)}$United Arab Emirates University, Al Ain; United Arab Emirates.\\
% ...
$^{123}$Homer L. Dodge Department of Physics and Astronomy, University of Oklahoma, Norman OK; United States of America.\\
% ...
$^{154}$Raymond and Beverly Sackler School of Physics and Astronomy, Tel Aviv University, Tel Aviv; Israel.\\
% ...
$^{a}$ Also Affiliated with an institute covered by a cooperation agreement with CERN.\\
$^{b}$ Also at An-Najah National University, Nablus; Palestine.\\
$^{c}$ Also at Borough of Manhattan Community College, City University of New York, New York NY; United States of America.\\
$^{d}$ Also at Center for Interdisciplinary Research and Innovation (CIRI-AUTH), Thessaloniki; Greece.\\
$^{e}$ Also at Centro Studi e Ricerche Enrico Fermi; Italy.\\
$^{f}$ Also at CERN, Geneva; Switzerland.\\
% ...

\end{flushleft}
